# Identifying Mirror Symmetry Density with Delay in Spiking Neural Networks

Jonathan K. George, Student *Member, IEEE*, Cesare Soci, Volker J. Sorger, *Senior Member, IEEE*

*Abstract* —**The ability to rapidly identify symmetry and anti-symmetry is an essential attribute of intelligence. Symmetry perception is a central process in human vision and may be key to human 3D visualization. While previous work in understanding neuron symmetry perception has concentrated on the neuron as an integrator, here we show how the coincidence detecting property of the spiking neuron can be used to reveal symmetry density in spatial data. We develop a method for synchronizing symmetry-identifying spiking artificial neural networks to enable layering and feedback in the network. We show a method for building a network capable of identifying symmetry density between sets of data and present a digital logic implementation demonstrating an 8x8 leaky-integrate-and-fire symmetry detector in a field programmable gate array. Our results show that the efficiencies of spiking neural networks can be harnessed to rapidly identify symmetry in spatial data with applications in image processing, 3D computer vision, and robotics.**

## I. Introduction

The human visual system is able to detect mirror symmetry rapidly [1]. Furthermore, symmetry detection in human vision is being hypothesized to be essential for 3D visualization [2]. While line integration has been proposed as a method for symmetry perception in spiking neural networks [3], a more fundamental role may be played by neural coincidence detection; in biology the spiking neural networks of the brain have been shown to be capable of both integration and coincidence detection [4], [5]. These spiking neurons have been observed to fire in a time-dependent manner forming strongly connected clusters [6] known as polychronous neural groups (PNGs) [5]. Both pattern recognition and computing have been achieved in artificial neural networks with polychronous behavior [7]. However, the connection between symmetry and coincident spiking polychronous neural networks has not been explored, nor has symmetry detection using only single layer delay in spiking artificial neural networks been demonstrated. Here we present a formal definition of geometric symmetry as the amplitude of a tensor space of the distribution of distance. We then show how a specific configuration of a spiking neural network can act on its inputs in a manner identical to a threshold applied to the tensor symmetry space, firing at the points of high mirror symmetry. As an example, we demonstrate a simple network both in software and in a Field Programmable Gate Array (FPGA) and validate symmetry-recognition capability of an artificial spiking neural network. The symmetry-associating behavior of spiking neural networks has immediate applications in image processing and is consistent with our intuition that the ability to identify symmetry is indeed supported by neural intelligence.

Jonathan K. George is with the Department of Electrical and Computer Engineering, George Washington University, Washington, D.C. USA (e-mail: jkg0@gwmail.gwu.edu).
Cesare Soci is with the Centre for Disruptive Photonic Technologies, Nanyang Technological University, Singapore, Singapore (email: csoci@ntu.edu.sg)
Volker J. Sorger is with the Department of Electrical and Computer Engineering, George Washington University, Washington, D.C. USA (e-mail: sorger@gwu.edu).

## II. Methods

Symmetry can be broadly defined as a self-similarity in logic or a dataset. Geometric symmetries are self-similarities in a spatial dataset. A geometric mirror symmetry is a symmetry across an axis. A geometric scaling symmetry is a symmetry of differing size. Similarly, a geometric rotational symmetry is a symmetry of rotation around a point. A geometric symmetry may include any combination of these transforms. In this paper, our focus is on geometric mirror symmetry which has been hypothesized to play an important role in human visual processing [2].

A geometric mirror symmetry is a reflection across some linear axis in the dataset. In the simplest case of a dataset consisting of two points in a space, the axis of mirror symmetry will be the line or plane that divides the two points. Each point on the line or plane will be equidistant from the two data points. Similarly, in a larger spatial dataset with a perfect mirror symmetry, the line or plane along the axis of mirror symmetry will have at least $N/2$ pairs of equidistant data points, where N is the number of data points in the dataset. That is, a line exists between each data point and its mirror and in the perfectly symmetric case all N/2 lines overlap along the axis of symmetry. In the perfectly symmetric dataset, the points along the line or plane of symmetry may have more than $N/2$ equidistant pairs, as other features in the dataset may have equidistant attributes, but they cannot have fewer than $N/2$ equidistant pairs. Finding the peaks in the distribution of equidistant points, then, is a good indicator of the presence of mirror symmetry.

### A. Definition of Mirror Symmetry Density

We define mirror symmetry density as the peaks of a continuous scalar field of the density of equidistant points using a histogram space. This is similar to other continuous definitions of spatial symmetry [8], [9]. Except in this case the problem is posed as a maximization of equidistant density on a histogram rather than simple density in [8] and a minimization of transform energy in [7]. While both perspectives of the problem result in the same solution (the equidistant point will be halfway between two points of the image and will have the minimum transform energy), viewing the problem in terms of maximization allows us to map the problem into a spiking neural network where thresholds are defined in terms of peak amplitude.

For a space to have symmetry density it must have some definition of distance. This distance function *f*, or metric, between two points ($A$ and $B$) must satisfy three conditions; First, it must be positive and equal for the same point.

$$f(a,b) \geq 0, f(a,b) = 0 \; if \; a = b \tag{1}$$

Second, it must be coordinate-symmetric, i.e. adhering to coordinate-reversal symmetry

$$f(a,b) = f(b,a) \tag{2}$$

Finally, it must satisfy the triangle inequality,

$$f(a,c) \leq f(a,b) + f(b,c) \tag{3}$$

which translates into the distance between two points being the shortest path. Meeting these criteria, a function is a metric. A common metric in the Euclidean space, for example, is the definition of distance in two-dimensional Cartesian coordinates:

$$\|B - A\| = \sqrt{(B_x - A_x)^2 + (B_y - A_y)^2} \tag{3}$$

With this definition of distance we define the set of symmetry points, $S = \{S_1, S_2, \ldots, S_n\}$, as the point equidistant between two points $A$ and $B$:

$$\exists X_j, X_k s.t. \|X_i - S_i\| = \|X_k - S_i\| \; for \; \forall \; S_i \in S, X_j \neq X_k \tag{4}$$

For the two points A and B, (4) defines the set of points forming a line equidistant between the two points A and B. For, three points A, B, and C, there are three lines of symmetry formed by each of the pairs (A, B), (B, C), and (A, C) as well as possibly a single point equidistant to all three points. As the number of input points increases, the number of symmetry lines increases with the number of unique pairs of input points.

### B. Algorithm

To understand the relationship of spiking neural networks with coincidence detection and to compare an artificial spiking neural network approach to a computational approach, it is useful to introduce an algorithmic method of ranking mirror symmetry. To develop a general algorithm to identify mirror symmetry, we first define a tensor field $\mathfrak{T}$ where the tensor value $\mathfrak{T}(x,y) = P(l)$ at each point in the coordinate space is a distribution of distances from that coordinate point to each input point ($l$). In this tensor field, high symmetry points will correspond to peaks in the distribution of distances. We define Algorithm 1 to generate a discrete representation of this field. With this algorithm symmetry points above a predefined threshold in the space can be identified. In $O(m * n)$ where $m$ is the number of points in our space $S$, and $n$ the number points in the set of input points in $N$. We note, that this algorithm can be applied iteratively where symmetries between input points (Fig. 1(a)) and other symmetry points create a hierarchy of symmetry points (Fig. 1(d)). The expected symmetry for the four inputs (green areas, Fig. 1(a)) result in the expected (1st order) symmetry shown in Figure 1(c).

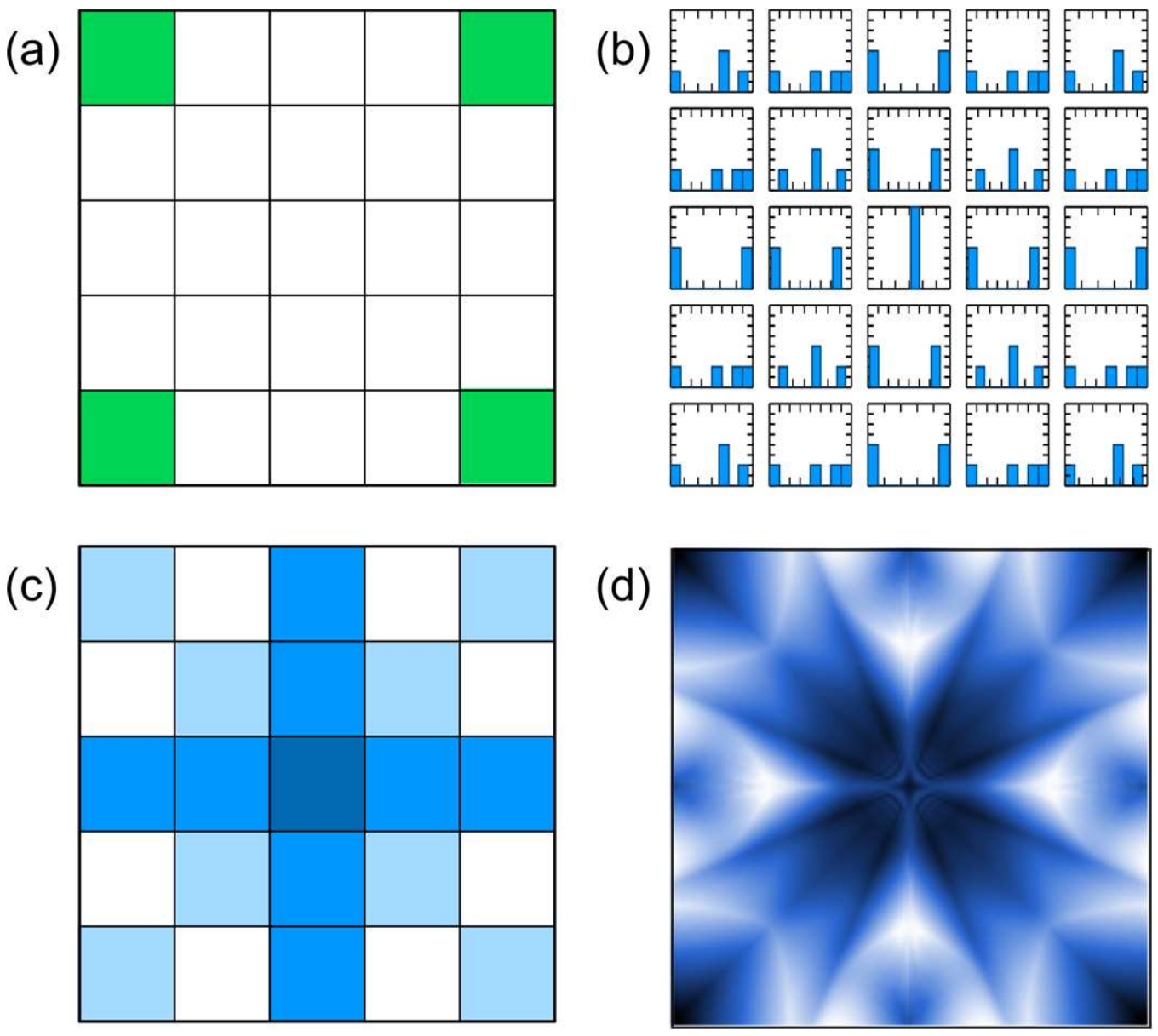

Fig. 1. The algorithm transforms the input data points (the four corners of a Cartesian space in this example) (a) into an array of histograms of distances (b) where each bin counts the distance from the point in space to each data point. The peaks of the set of histograms are used to create a density of symmetry (c) overlaying the original space. Applying the algorithm repeatedly with feedback in a high-resolution space produces a fractal-like pattern (d) of hierarchical symmetry.

### C. Symmetry Density in Spiking Neural Networks

Next, we show how a spiking network enables coincidence detection, which together with the aforementioned symmetry Algorithm 1 allows for a neural network implementation of symmetry detection.

**Algorithm 1** Mirror Symmetry

```
Input: P_{x,y}: spatial data, threshold: ratio from 0 to 1
Output: S_{x,y}: spatial boolean symmetry
  Initialization :
  S_{x,y} := 0
  T_{x,y} := 0
  maxsym := 0
  Tensor Process
  for s_{x,y} in S do
    for p_{x,y} in P do
      d := distance(s_{x,y}, p_{x,y})
      T_{x,y}[d] := T_{x,y}[d] + 1
      if T_{x,y}[d] > maxsym then
        maxsym := T_{x,y}[d]
      end if
    end for
  end for
  Symmetry threshold
  for t_{x,y} in T do
    for d in t_{x,y} do
      if t_{x,y}[d] > maxsym * threshold then
        S_{x,y} := 1
      end if
    end for
  end for
  return S
```

A spiking neural network is a type of artificial neural network that models the spiking observed in the neuron cells of the brain, in this sense it is Neuromorphic, taking on the form of a neuron. By concentrating its energy into a short time span the spike is an efficient encoding method in a noisy environment [10].

A simple type of spiking neural network is the Leaky Integrate and Fire (LIF) model [11]. In this model each neuron integrates each of its inputs in time while simultaneously leaking from the accumulator. When the accumulator passes a threshold level it fires, generating a signal spike. The leak creates a temporal dependence on the past, thus adding memory to the neuron.

The LIF model can be formulated mathematically [11] as

$$u(t) = RI(t) - \tau_m \frac{du}{dt} \tag{5}$$

where the voltage $u$ is a function of current with a leaky term $\tau_m \frac{du}{dt}$ that depends on the change in voltage with time, and $\tau_m$ is the relaxation time constant of the signal leak to reach threshold.

To understand how spatial symmetry affects the result of a LIF neural network, consider a simple neural network consisting of two input neurons A and B connected to a single output neuron C (Fig. 2). This three-neuron system acts as a coincidence detector; for a spike to propagate from neuron A to neuron C some propagation time $\Delta t_A$ must pass. Similarly, for a spike to propagate from neuron B to neuron C some propagation time $\Delta t_B$ must pass. If neuron C is assigned a threshold that is just higher than the individual spikes output from either A or B, then the neuron C will not fire unless both spikes arrive time-synchronized at C within a temporal window before the leaky time constant $\tau_m$ reduced the power below threshold at C. Assuming a constant propagation speed across the system, if both neurons A and B are fired in synchrony with one another, the spike arrival times $\Delta t_A$ and $\Delta t_B$ will be dependent only on propagation distances $d_A$ and $d_B$. If the two distances are equal the propagation times will be equal and the two spikes will arrive together at C, pushing it over threshold and causing it to fire. On the other hand, if the distances are not equal, the propagations times will be different, the two spikes will arrive separately at C and the neuron will not fire. In this way, synchronized neurons in spiking neural networks act as the distance-relating components in the distribution elements of our tensor space $\mathfrak{T}$, and the threshold at neuron C acts as the second half of our symmetry algorithm, selecting the peaks out of the distance distributions in $\mathfrak{T}$. This method effectively, enables the neuron to act as a coincidence detector (Fig. 2).

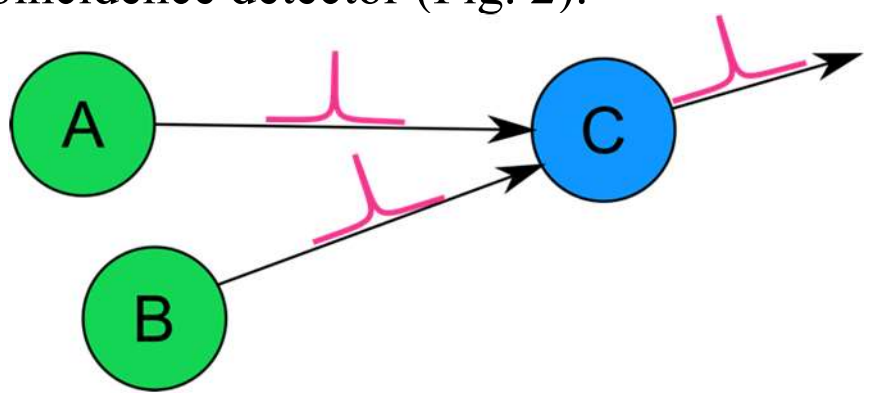

Fig. 2. When neuron A and neuron B fire simultaneously, and if both neurons are equidistant from neuron C, then the pulse propagating from A to C will arrive at C at the same time as the pulse propagating from B to C. The threshold at C is set such that it is higher than the amplitude of either individual pulse. Neuron C will fire if, and only if, both pulses arrive time-synchronized. In this way neuron C acts as a coincidence detector.

Extending upon the three-neuron system, this concept of coincidence detection via delay can be brought to the entire space (Fig. 3). Each neuron of the input layer is connected to each neuron of the output layer (*N²-2N* connectivity) with a delay proportional to the distance between the two points in the space. For example, in a Cartesian space if input neuron 1 is point x = 0, y = 0, and output neuron 3 is the point x = 5, y = 4, the delay would be proportional to $\sqrt{25+16} \approx 6.4$. As stated before, the coincidence detecting property of the spiking neuron is selecting equidistant points.

.

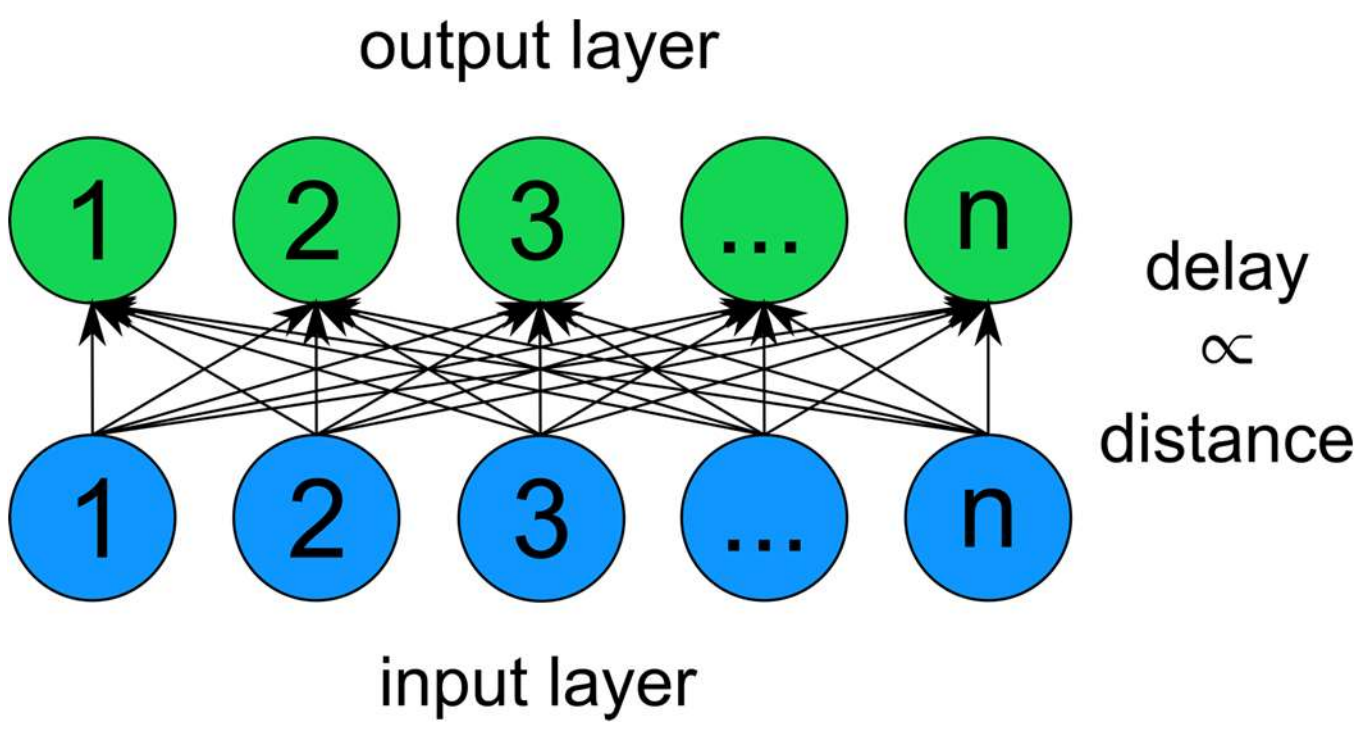

Fig. 3. Each input layer neuron is connected to every output layer neuron with a delay proportional to the distance between the input and output points in Euclidean space. If all activated input neurons fire simultaneously, and only one pulse is allowed per cycle, the output node with highest symmetry will be the output node with the greatest number of pulses arriving simultaneously.

The temporal response (speed) of such a system in the ultimate physical limit is defined by the sum of the input layer firing delay, the pulse propagation delay, and the delay of the threshold layer. Assuming speed of light for propagation delay, the resolution of the device is set by its ability to distinguish individual pulses within the time of propagation. The worst case is the time delay of the smallest distance, given the shortest distance between individual data points in the space, i.e. the pitch of the array. Then the threshold layer must switch with at least $\Delta t = d_p/c$ where $d_p$ is the pitch distance. From this we can state a lower bound for the average energy consumption of the threshold layer in the ultimate speed limit [12] (Eq. 6).

$$E \geq \frac{\pi\hbar}{2\Delta t} = \frac{\pi\hbar c}{2d_p} \tag{6}$$

### D. Layering

Deep Neural Networks (DNNs) are inherently multilayered. The multilayered architecture gives DNNs the potential to create higher levels of abstraction than single layer networks. It is reasonable to ask if the geometric-symmetry identifying neural networks discussed here are able to support layering.

If we naively begin with the network of Fig. 3 and add a third layer on top of the original output layer, we quickly discover that the layering cannot be supported with the network as previously presented. While the pattern of symmetry will appear at the middle layer, each coincident set of pulses will arrive at different times. In order for geometric symmetry to be identified by coincidence detection in the top layer, input pulses (i.e. output from the middle layer) must either all start at the same time or, if initiated with time differences, these individual delays must be proportionally shorter than the respective connection delays. If the time differences were constant we could remove time from the connection delay between the middle layer and the new output layer. However, the time differences are not predictable and are dependent on the symmetry in the data from the input layer. If we are to detect geometric symmetry in a multilayered network, we must adjust the network to account for the varying arrival times.

The LIF spiking neuron can act as a memory for synchronization, much like a register in digital logic. To see this behavior we imagine a set of pulses independently arriving at a set of slow leaking LIF neurons such that some of the neurons receive a pulse and some do not. The pulses may represent the 1s and 0s of the bits to be stored in the memory. Now, with the proper choice of threshold, the bit will be stored until either the neuron leaks away all of the energy received by the pulse or another pulse arrives, pushing the neuron over the threshold. If this second set of pulses arrives at every neuron at the same time, the neurons will act as a synchronization stage, collecting pulses from their input and waiting until activated to simultaneously release the stored pulses to their outputs.

We can apply this concept of synchronization to the multilayered geometric symmetry-identifying network by adding a synchronization layer above the original output layer (Fig. 4). This synchronization layer will have a slower leak rate than the layer below it so that it stores the symmetry points generated by the geometric symmetry detecting network below and, when a timing pulse arrives, release them at once to the symmetry detecting network above. In this way we can build a network to detect hierarchical geometric symmetry, i.e. the symmetry points of symmetry points.

Using the same process of synchronization we can also feed the output layer back to the input layer. The feedback allows the network to act on the symmetry points as well as the original data. The feedback output will differ from the multilayered network, as now original data will be compared to the generated symmetry points, whereas in the multilayered network each layer only computes the geometric symmetry of the layer below it. This feedback will create a dynamic system with results similar to repeated application of the symmetry algorithm presented earlier (Fig. 1(d)).

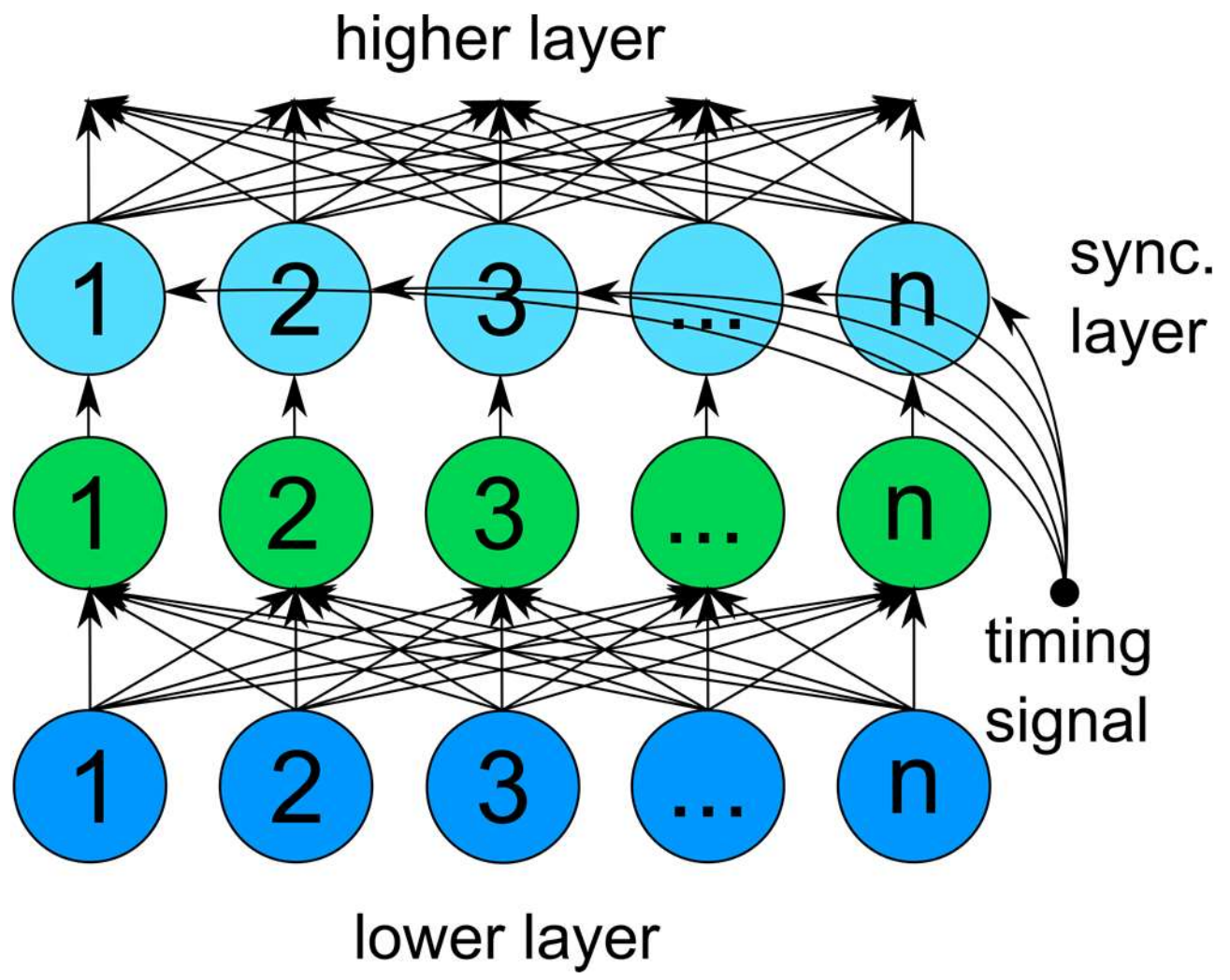

Fig. 4. To detect hierarchical geometric symmetry a synchronization layer is added above the original output layer. The synchronization layer has a slower integration time than the original output layer, storing the pulses until a timing signal releases them to the next higher layer simultaneously. For feedback, the output of the synchronization layer is connected back to an input layer. See text for details of operation.

### E. Sets

One of the primary applications of artificial neural networks is classification. Here the neural network decides the class of new input data based on prior training. In classification problems, it is often useful to compare sets of data. It may be useful in this context to find the geometric symmetry between two different sets of spatial data. If we attempt to add two sets of data at the input layer, one for class A and one for class B, we will create a network that finds geometric symmetry between data in A and B but it will also find geometric symmetry from points in A to points in A and from points in B to points in B. To create a network that only detects symmetry between two sets of data we must amend the aforementioned architecture, as discussed next.

Each coincidence detector must now operate only on inputs from the two different sets and not on two inputs originating from the same set. One way to achieve this constraint is to create a coincidence detector for every pair of equidistant inputs arriving from the two input sets at every point in the output space. While this implementation would achieve the result we seek, it would also require an impractical number of coincident detecting neurons.

An alternative implementation separates the two sets through pseudorandom delay coding. First, the time resolution of the output layer is increased by slightly more than three-times the edge length of the array by increasing the leak rate of each neuron, the time resolution of the delay, and the time resolution of the output detector. Next, a time-dispersing pseudorandom code is applied to the connection delays out of the input layer of set A. This time dispersing code separates the pulses of the set such that no two pulses from set A will arrive the output layer within the integration time of the output layer, preventing intra-set symmetry detection. Finally, the identical pseudorandom code is applied to the output delays out of the input layer of set B. Now, the network will detect inter-set symmetry of data points from the two data sets without detecting intra-set symmetry points, and the number of neurons only needs to be increased by the number of data points in the additional set. This principle can be applied in the same manner to any number of data sets.

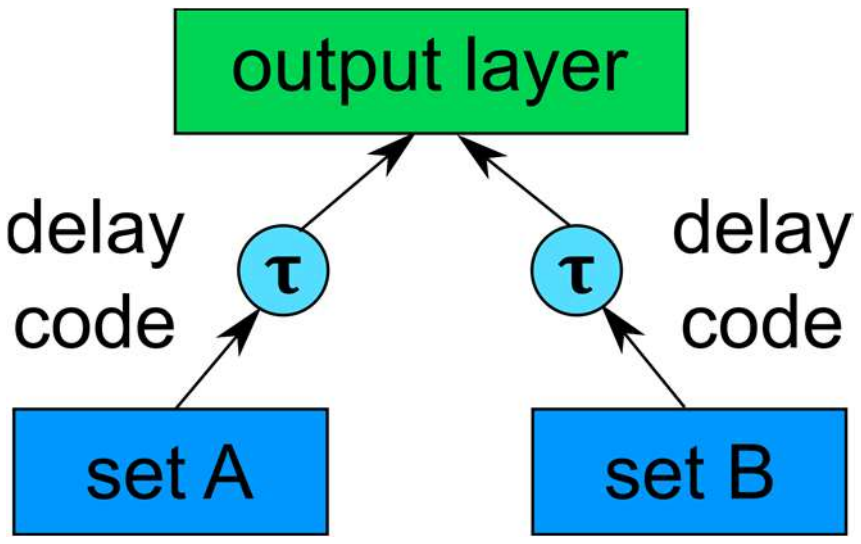

Fig. 5. An identical time-dispersing pseudorandom delay code τ is applied to add additional delay to both input sets A and B. This delay prevents intra-set symmetry from being detected at the output layer by preventing equidistant pulses originating from the same set from arriving simultaneously within the integration time of the output layer, while allowing inter-set symmetry detection by preserving coincidence detection for equidistant points between the two input sets.

The time dispersing code must separate all equidistant points connecting into each node in the output layer. The center output node always has the highest number of equidistant inputs. For example the center point of a 5x5 array has 12 equidistant inputs when distance is rounded to the nearest integer. The maximum number of equidistant inputs increases linearly with approximately 3 times the edge length of the array. Each of these equidistant points must have pulses arriving at different times to prohibit intra-set symmetry detection. In the case of the 5x5 array this would require 12-fold increase in the time resolution compared to the intra-set allowing case. The time dispersing implementation trades a significant increase in required neurons for a significant increase in required time resolution.

### *F. Metrics*

The concept of symmetry density is not limited to Cartesian coordinates or Euclidean distance. Other measures of distance can form varying types of equidistant symmetry. In the implementation of the algorithm in software, the Manhattan distance may be appealing as it can simplify the calculation of distance by eliminating the squares and square root of the Euclidean distance formula. However, this simplification has dramatic effects on the distribution of symmetry. Given points A and B in Manhattan space, a new equidistant point C can be found by moving B diagonally. In Euclidian space the equidistant point C is found by moving B circularly around point A. This has the effect of emphasizing horizontal, vertical, and diagonal lines in Manhattan space (Fig. 6).

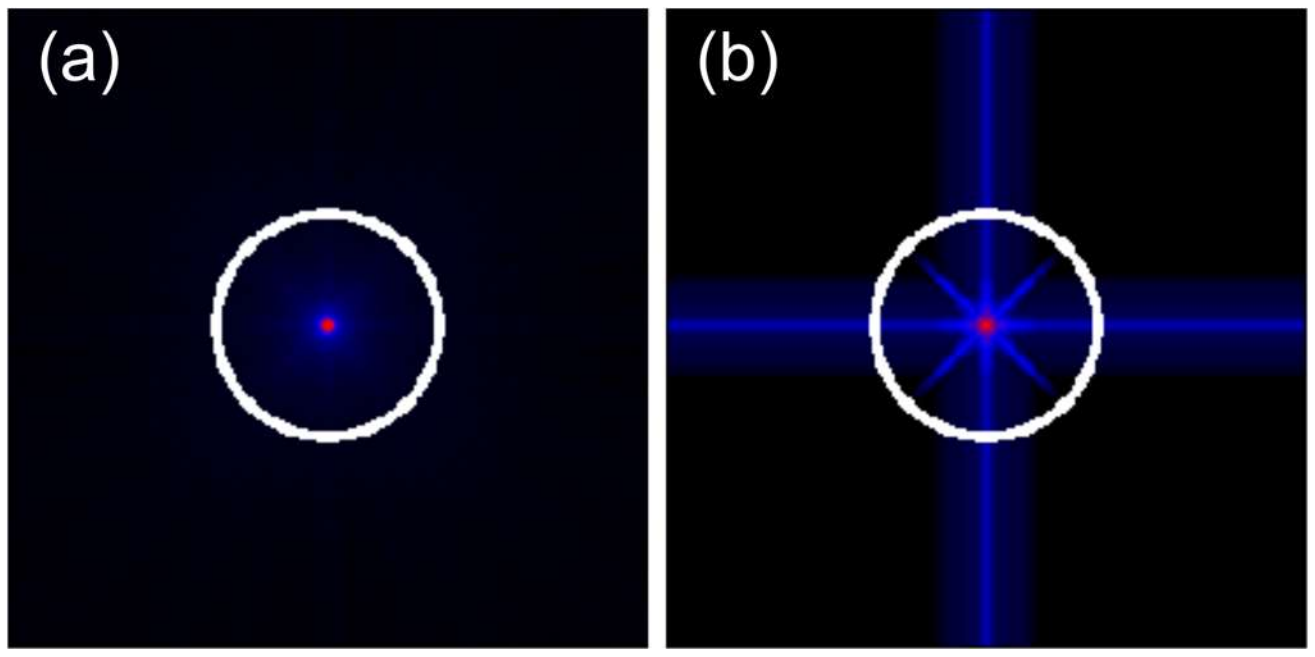

Fig. 6. A comparison of the magnitude of symmetry density of a circle in both Euclidean space (a) and Manhattan space (b) shows the Manhattan space favoring coordinate-parallel and coordinate-diagonal lines, as expected.

### *G. Noise*

To withstand noise, implementations of the symmetry density algorithm will need to set a threshold to determine which points are included in the dataset or weight the distribution with the value of the pixels. In the threshold case, the data point is either included or excluded from the distribution of distances based on whether the value of the pixel passes some threshold value. In a spiking neural network this is the activation threshold of the neuron receiving the input signal. The threshold can be set to be the mean or some standard deviation above or below the mean. In the weighting case the average value, or minimum value of the two equidistant pixels is applied as a weight to the value of the symmetry density in the same way that mass is applied as a weight in finding the center of mass of an object. This result can also be seen in long averages of a spiking neural network where the repetition of pulses is proportional to the amplitude of the input signal.

Adding noise to the input causes the addition of noise in the distance distribution, raising the noise floor of the distribution. This noise decreases the separation between the peak of the distribution and the noise floor. Once the original peak is no longer detectable in the distribution, the output point in the symmetry density will become incorrect. The strength of noise where this occurs is dependent on the Signal to Noise Ratio (SNR) of the original symmetry density.

Gaussian noise is independent of the signal and will thus have a flat distance distribution. Adding Gaussian noise to the input will have a proportional effect on the SNR of the symmetry density. Any addition of noise in dB can simply be subtracted from the SNR of the original symmetry density to determine the resulting SNR.

Gaussian noise is, however, unrealistic in most imaging systems where physically counting photons produces Poisson distributed noise. Unlike Gaussian noise, Poisson noise is dependent on the signal amplitude (i.e. number of photons). In this case the addition of noise will not be distributed evenly in space but will follow the inverse of the signal. That is, pixels receiving fewer photons will have greater noise than pixels receiving a greater number of photons. As the image becomes less and less exposed, the separation between the light and dark pixels diminishes until objects are no longer distinguishable. In the Poisson case images with both light and dark pixels experience both small and large noise sources simultaneously. In a thresholding implementation, on a sufficiently exposed image, the dark pixels are significantly beneath the average brightness value and will be excluded by the threshold. In this case, the symmetry density is only effected by noise of the brightest pixels, thus favoring low-noise data. Only when the image exposure is reduced to the point that distribution of dark pixels and bright pixels begins to overlap will the symmetry density be affected.

## III. Results

### A. Implementing LIF with Digital Logic

Each neuron is represented as a leaky accumulator. At each clock cycle all of the neuron inputs are added to a value stored in an accumulation register, the result of the summation minus the leak value are then stored back in the accumulation register. If the accumulation register surpasses the value of a fixed threshold, or, alternatively as a configurable threshold stored in a threshold register, the accumulator is reset to zero and a value appears at the neuron's output. If the accumulation value does not surpass the threshold, zero appears at the neuron's output.

Neuron-to-neuron connection delay is represented either as queue, where the outputs are placed into a First-In-First-Out (FIFO) queue, or as a pulse value and countdown register. When delay is represented as a queue, at each clock cycle a single value is added from the output neuron to each output connection's FIFO and a single output is removed from each FIFO at the output neuron. The length of each FIFO is proportional to the delay being represented. When delay is represented as a countdown register, at each clock cycle if the neuron's output value is positive a countdown register is initialized with a count proportional to the represented connection delay and a value of the neuron's output value. At each clock cycle each countdown register is decremented. When a countdown register reaches zero, its pulse value register is placed at the input of the output neuron.

### B. FPGA Implementation

To demonstrate spatial symmetry recognition via coincidence detection of LIF networks on actual hardware, we implemented a simple LIF spiking neural network on a Xilinx Zynq™ FPGA. Our LIF neural network consists of an 8x8 input array connected to an 8x8 neuron output array. Each output neuron is connected to every input by a shift register of length proportional to the Manhattan distance from the point ($O_x$,$O_y$) at the output to the point ($I_x$, $I_y$) at the corresponding input. This results in 4,096 shift registers with a maximum length of 16. Each output neuron consists of a two-stage adder followed by an accumulation register. Each adder includes a configurable constant leaky term that subtracted the configured leak from the accumulation register at each time interval. Each accumulator is connected to a threshold level. If the accumulator passes the threshold, a second single-bit register is set to 1 to indicate firing of the output neuron. The latency from input to output in this implementation is proportional to the length of the longest shift register, 16 in this case, plus the accumulation time, 2 in this case, for a total of 18 clock cycles. The network is clocked at a constant clock speed of 50 MHz for approximately 2.8 MHz of 8x8 symmetry operations.

The output of the symmetry LIF neural network was recorded over time for the elementary case of a line between two points (Fig. 7). Our results confirm that spiking LIF neural networks indeed act as symmetry detectors, highlighting equidistant points in spatial data.

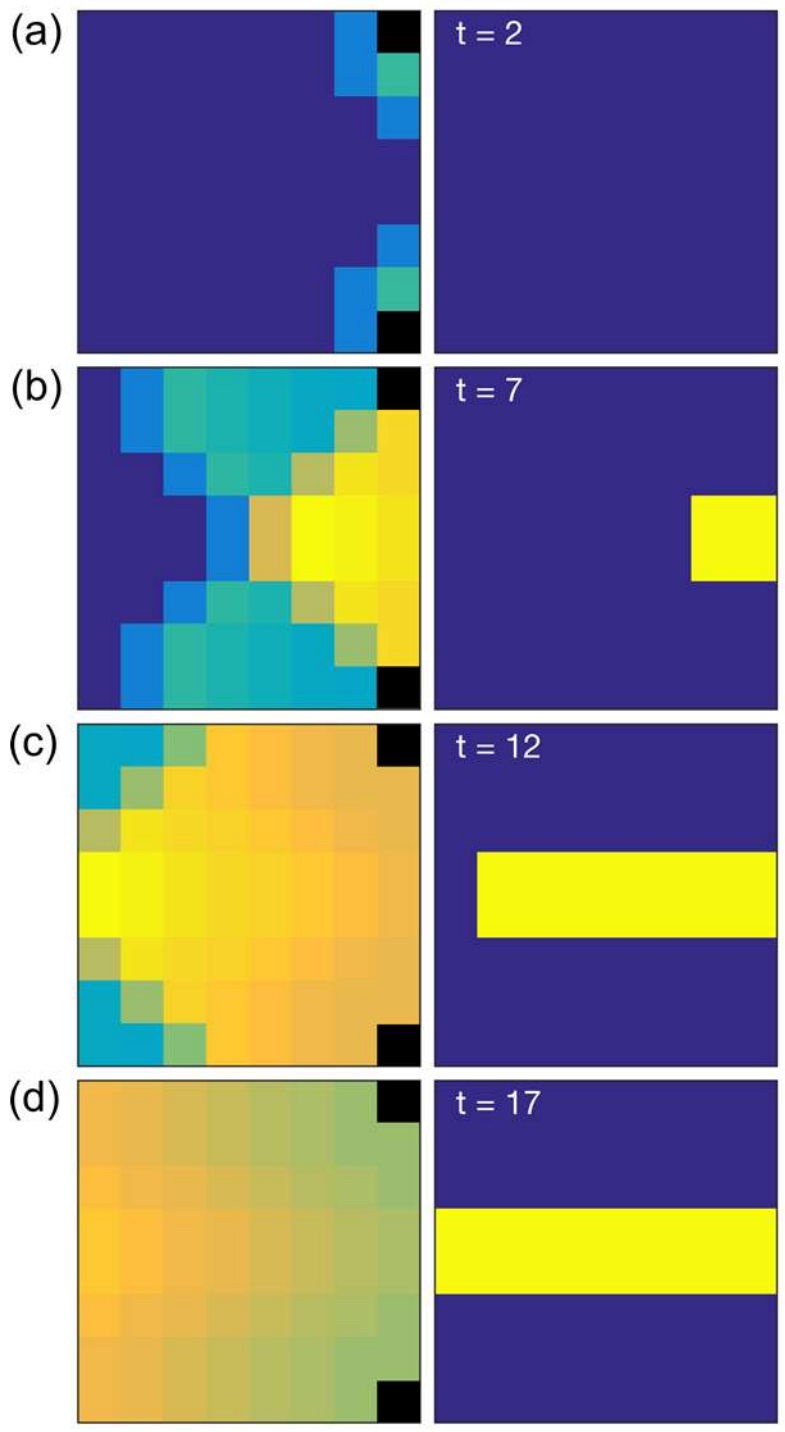

Fig. 7. Time-evolution output of the symmetry detection algorithm using the coincidence detection of LIF neural network implemented in a Xilinx Zynq™ FPGA. From two-point source data (black dots in (a) of the 8x8 array), (a) t = 2 clock cycles into the 18 clock cycle period with raw output of the accumulation register array (left) and threshold register array (right), shows identification of the line of symmetry points, (b) at t = 7 threshold begins to be reached at the center points on the right (c) at t = 12 peak continues to propagate to the left (d) peak exits the left of the array finishing a trace of a two point line between the original data points.

### C. Analysis of the MNIST Dataset

One of the primary applications of image processing and neural networks is image classification. The objective of image classification is to identify the set to which an unlabeled image belongs. We hypothesize that if symmetry is central to the human visual system, that mirror symmetry density may prove useful in classifying human generated symbols, such as handwriting and printed text. To evaluate effectiveness of symmetry density in image clustering we applied the symmetry density algorithm to 55,000 images of the MNIST (Modified National Institute of Standards and Technology database) dataset. This database is a database of handwritten digits used for training various image processing systems.

In neural network image classification, the neural network may be trained on a labeled training dataset and then tested against a second labeled evaluation dataset.

### *D. Additive Noise and the MNIST Dataset*

To evaluate the effect of noise on the mirror symmetry density, we processed 65,000 images of the MNIST handwritten number dataset with additive Gaussian noise through Algorithm I. Gaussian additive noise with a variance from 0 to 1 in increments of 0.1 was applied to the dataset. The data points greater than the mean of the noised data were included and data points less than the mean were set aside. The Mean Square Error (MSE) between the result of processing the thresholding noise data by Algorithm I and the original dataset were computed (Fig. 8).

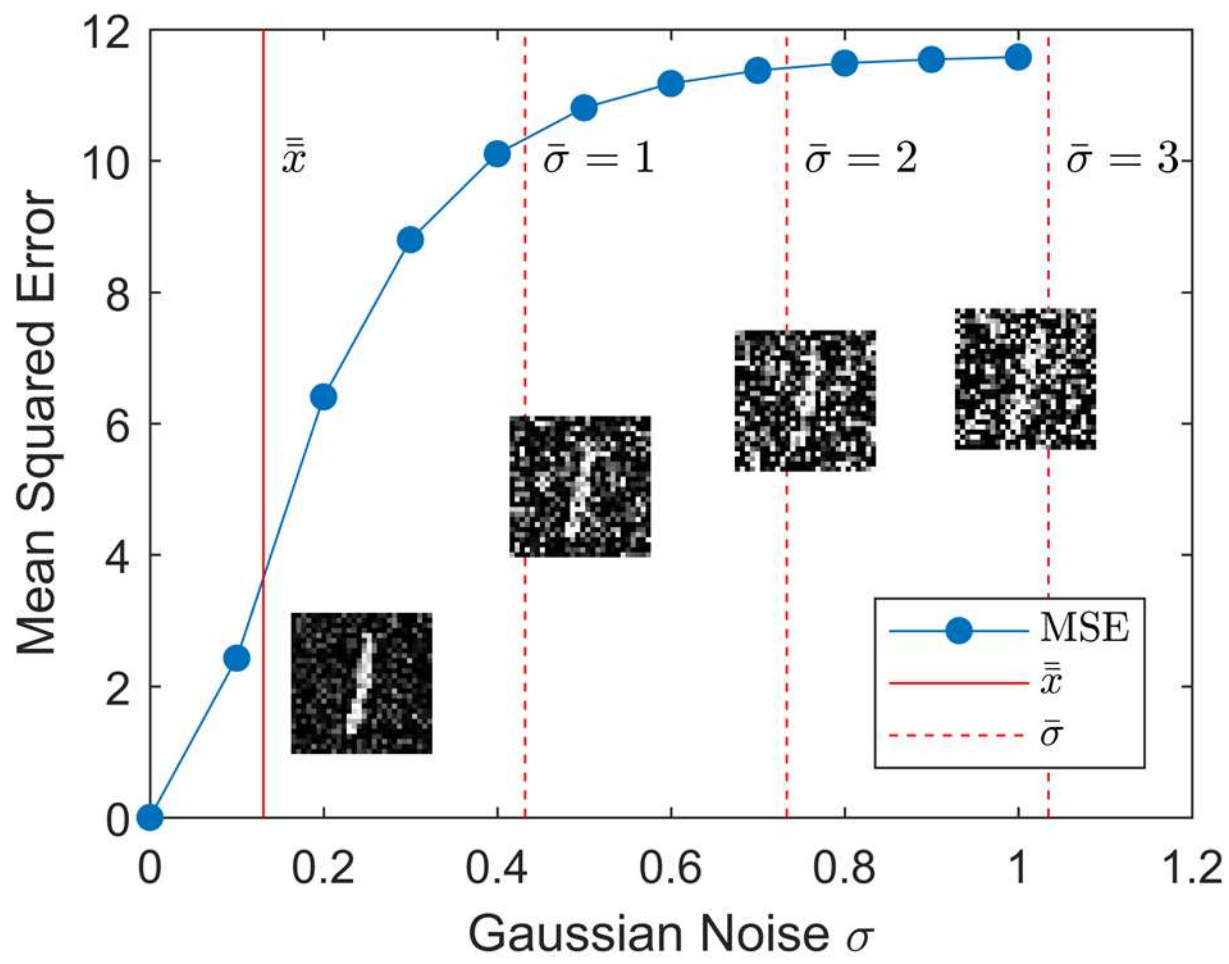

Fig. 8. Gaussian additive noise with $\sigma$ swept over 11 points from 0 to 1 applied to 65,000 images of the MNIST handwriting dataset shows that when using a threshold of the mean of the noised data, $\bar{x}$, the MSE increases asymptotically to saturate soon after the noise variance passes the mean first standard deviation, $\bar{\sigma}$. This is seen in the four sample images (inset) with increasing $\sigma$ of 0.25, 0.5, 0.75, and 1 respectively. This is the expected result with the Gaussian spatial noise being independent of the input data and having proportional effect on SNR of the distribution of distance.

## IV. Conclusion

In conclusion, we have presented a novel algorithm for finding a scalar field representing the symmetry of points in a multi-dimensional space. We have shown how time synchronization in the input values of spiking neural networks, with the appropriate choice of threshold and spike period, results in the identification of output neurons along points of high symmetry density to the network inputs. We have demonstrated an implementation of the symmetry selective LIF neural network in common hardware with a high speed, 2.8 MHz identification of symmetry points in an 8x8 Manhattan metric space. Our results show that utilizing only the delay and coincidence detecting properties of a single layer of neurons in spiking neural networks naturally lead to effective symmetry identification. A greater understanding of symmetry perception in artificial intelligences will lead to systems with more effective pattern visualization, compression, and goal setting processes. Future research should focus on optical implementations of the presented findings a) to harness the information parallelism of bosonic photons, b) to chaptalize on the high energy efficiency of photonic and nanophotonic optoelectronics which only require the micrometer-small active devices to the electrically addressed enabling 100's of atto-Joule efficient active optoelectronic devices, and (c) to enable high-data throughput links and platforms [13]-[22].

## Acknowledgements

V.S. is supported by Air Force Office of Scientific Research-Young Investigator Program under grant FA9550-14-1-0215, and under grant FA9550-15-1-0447.

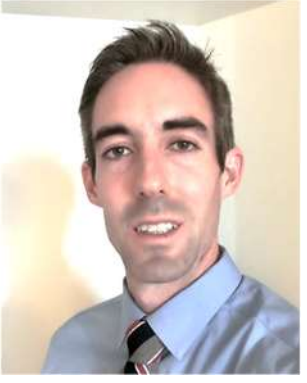

**Jonathan K. George** received the BS degree in computer science and applied mathematics from the University of Colorado at Colorado Springs in 2005 and the MS degree in electrical engineering in 2015 and is currently pursuing the PhD in electrical engineering at the George Washington University, Washington DC. He is a Senior Principal Engineer at Orbital ATK in Dulles, Virginia. His research interests include artificial intelligence, optical computing, and nonlinear optics. He is a student member of IEEE, OSA, and SPIE.

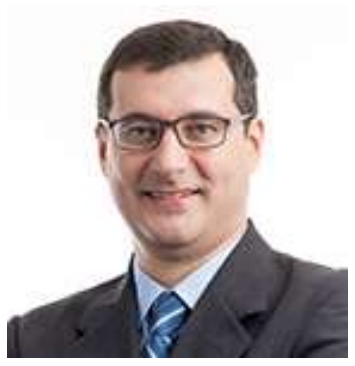

**Cesare Soci** is an associate professor at Nanyang Technological University, Singapore and the deputy director of the Centre for Disruptive Photonic Technologies. Soci received Laurea and Ph.D. degrees in Physics from the University of Pavia, in 2000 and 2005. He was a postdoctoral researcher from 2005 to 2006 at the Center for Polymers and Organic Solids of the University of California, Santa Barbara, and from 2006 to 2009 at the Electrical and Computer Engineering Department of the University of California, San Diego. He joined the Nanyang Technological University (NTU) in 2009, where he holds a joint appointment between the Schools of Physical and Mathematical Sciences (SPMS) and Electrical and Electronic Engineering (EEE).

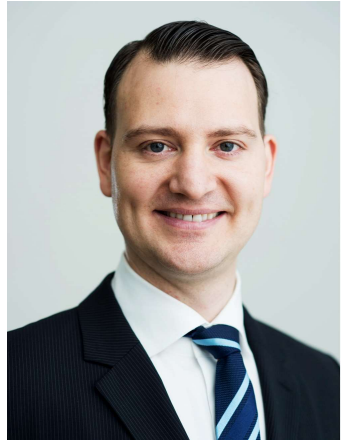

**Volker J. Sorger** is an associate professor in the Department of Electrical and Computer Engineering, and the director of the Orthogonal Physics Enabled Nanophotonics (OPEN) Labs at the George Washington University. He received his PhD from the University of California Berkeley. His research areas include opto-electronic devices, optical information processing, and internet-of-things technologies. Dr. Sorger received multiple awards such as the Hegarty Innovation Prize, the Young Career Award at GWU, the AFOSR Young Investigator Award, and the best annual paper award by the National Academy of Sciences. Dr. Sorger is the executive chair for the technical groups of OSA. He serves at the board of meetings for both OSA and SPIE. He is the editor-in-chief for the journal Nanophotonics, and senior member of IEEE, OSA, and SPIE.